\documentclass[10pt,twocolumn,letterpaper]{article}

\usepackage[pagenumbers]{cvpr} % To force page numbers, e.g. for an arXiv version

\usepackage{graphicx}
\usepackage{amsmath}
\usepackage{amssymb}
\usepackage{booktabs}
\usepackage{framed,multirow}

\usepackage[pagebackref,breaklinks,colorlinks]{hyperref}

\usepackage[capitalize]{cleveref}
\crefname{section}{Sec.}{Secs.}
\Crefname{section}{Section}{Sections}
\Crefname{table}{Table}{Tables}
\crefname{table}{Tab.}{Tabs.}

\begin{document}

\title{3D Intracranial Aneurysm Classification and Segmentation via Unsupervised Dual-branch Learning}

\author{Di Shao\\
Deakin University\\
75 Pigdons Rd, Waurn Ponds, 3216, Australia\\
{\tt\small shaod@deakin.edu.au}
\and
Xuequan Lu\\
Deakin University\\
75 Pigdons Rd, Waurn Ponds, 3216, Australia\\
{\tt\small xuequan.lu@deakin.edu.au}
\and
Xiao Liu\\
Deakin University\\
75 Pigdons Rd, Waurn Ponds, 3216, Australia\\
{\tt\small xiao.liu@deakin.edu.au}
}
\maketitle

%%%%%%%%% ABSTRACT
\begin{abstract}
Intracranial aneurysms are common nowadays and how to detect them intelligently is of great significance in digital health. While most existing deep learning research focused on medical images in a supervised way, we introduce an unsupervised method for the detection of intracranial aneurysms based on 3D point cloud data. In particular, our method consists of two stages: unsupervised pre-training and downstream tasks. As for the former, the main idea is to pair each point cloud with its jittered counterpart and maximise their correspondence. Then we design a dual-branch contrastive network with an encoder for each branch and a subsequent common projection head. As for the latter, we design simple networks for supervised classification and segmentation training. Experiments on the public dataset (IntrA) show that our unsupervised method achieves comparable or even better performance than some state-of-the-art supervised techniques, and it is most prominent in the detection of aneurysmal vessels. Experiments on the ModelNet40 also show  that our method achieves the accuracy of 90.79\% which outperforms existing state-of-the-art unsupervised models.
\end{abstract}

%%%%%%%%% BODY TEXT

\section{Introduction}\label{sec:1}

Intracranial aneurysms can result in a high rate of mortality, and their classification and segmentation are of great significance. Existing research mainly focused on image data which involve regular pixels \cite{nakao2018deep,stember2019convolutional,ueda2019deep,sichtermann2019deep,shi2020clinically,joo2020deep}. While 3D geometric data such as point clouds can depict more useful information, the research on analysing intracranial aneurysms using point cloud data has been very sparsely exploited to date. Thanks to \cite{yang2020intra}, a point cloud dataset including aneurysmal segments and healthy vessel segments has been published. They have conducted a benchmark using state-of-the-art point-based networks that can directly consume 3D points instead of 2D pixels.

There are many networks available for consuming point cloud data, for example, PointNet \cite{qi2017pointnet}, PointNet++ \cite{qi2017pointnet++}, SpiderCNN \cite{xu2018spidercnn}, PointCNN \cite{li2018pointcnn}, SO-Net \cite{li2018so-net} and DGCNN \cite{wang2019DGCNN}. PointNet is a seminal method for taking 3D points as input and used for 3D point cloud classification and segmentation. Later on, other point-based methods have been proposed to improve the performance. Since they are all supervised learning methods, annotated data are required for training. However, annotation often requires experts and significant amounts of time, especially for large datasets and medical data. 

With the above analysis in mind, we design an unsupervised representation learning method that consumes point clouds of vessel segments. The contrastive learning concept inspires our method. In particular, we first generate a pair of augmented sample of the original point cloud which should have a distinctly difference. Next, we design a dual-branch contrastive network with an encoder for each branch and a follow-up common projection head to facilitate the unsupervised training with a contrastive loss. As for the downstream tasks, we first use the unsupervised trained model to output the representations. Then, we design simple networks and train it by taking the representations as input to classify or segment intracranial aneurysms. Note that we design two unsupervised networks and two corresponding downstream networks to fulfill two different tasks (i.e. classification and segmentation). Supervised methods often need a large scale of labelled data for achieving satisfactory performance. Compared with them, our method does not require labels in unsupervised training, and it can utilise a small scale of labelled data for downstream training. In summary, our contributions in this paper include:
\begin{itemize}
    \item We propose a simple yet effective method for unsupervised representation learning on 3D point clouds of vessel segments.

    \item We invent a useful augmentation method for generating pairs of each vessel segment.
    
    \item We propose a dual-branch contrastive network with an encoder for each branch.

    \item We conduct comprehensive experiments and compare with state-of-the-art point-based techniques to demonstrate the superior performance of our method. 
\end{itemize}

\section{Related work}\label{sec:2}
\subsection{Deep Learning on Intracranial Aneurysms}\label{subsec:2.1}
Intracranial aneurysms are associated with a high mortality rate. Therefore, the detection of intracranial aneurysms is crucial for human health. Traditional methods rely greatly on prior knowledge, which is often inferior to deep learning in terms of capability and accuracy. Due to the excellent performance of deep learning in processing medical images, there are many deep learning methods to detect intracranial aneurysms \cite{shi2020artificial}.  \cite{nakao2018deep} proposed a convolutional neural network-based detection system. The system used a 6-layer (CNN) and maximum intensity projection (MIP) algorithm based on the MRA images. This method can achieve almost 100\% accuracy for detecting aneurysms greater than 7 mm in diameter. However, it was less sensitive for small vascular aneurysms. To improve this, \cite{stember2019convolutional} used the full U-net convolution architecture to predict aneurysm size based on the detection.  \cite{ueda2019deep} applied the ResNet-18 network to the MRA images and performed a secondary evaluation on the already detected image data to enhance the detection sensitivity. To better segment the shape of intracranial aneurysms,  \cite{sichtermann2019deep} utilized DeepMedic \cite{kamnitsas2016deepmedic} with 2-pathway architecture and 11-layer convolution to segment intracranial aneurysms from the MRA images on the basis of detection. The above method for detecting intracranial aneurysms used data that are stacked with 2D images. To sum up, nearly all works focused on dealing with medical images rather than 3D geometry like point clouds.

\subsection{Point-Based Networks}
\label{subsec:2.2}
Neural network models for the classification and segmentation of 3D point cloud data have achieved noticeable successes. \cite{qi2017pointnet} proposed PointNet to directly process point sets. To obtain permutation invariance and transformation invariance of point clouds, PointNet used the symmetric function and T-net to design the network. It had good results for global features extraction of point clouds. However, it ignored the geometric relationship among points and limited the extraction of local features. To address this issue, \cite{qi2017pointnet++} proposed PointNet++ using a hierarchical neural network. It used the point sampling and grouping strategy to extract local features of point clouds. However, PointNet++ did not reveal the spatial distribution of the input point cloud. SO-Net \cite{li2018so-net} constructed the Self-Organizing Map (SOM) \cite{kohonen1990self} to model the spatial distribution of the input point cloud. It allowed SO-Net to adjust the receptive field overlap and performed hierarchical feature extraction. Unlike SO-Net with adjusting the perceptual field of the hierarchical network, PointCNN \cite{li2018pointcnn} proposed the $\chi$-transformation to process the point cloud data so that the point cloud data can be weighted or permuted. Thus, it improved the extraction of local features.
In addition, SpiderCNN \cite{xu2018spidercnn} proposed SpiderConv, i.e. parameterized convolutional Filters, to implement convolutional operations on disordered point clouds. DGCNN \cite{wang2019DGCNN} proposed a convolutional-like operation by constructing local neighbourhood graphs and applying convolutional operations on the edges. It connected adjacent point pairs to exploit the local geometric structure. In addition to common tasks like classification and segmentation, point based networks have also been developed to address other tasks \cite{zhang2020pointfilter, wang2021deep}. In summary, there has been great progress in analyzing point cloud data in a supervised manner.

\subsection{Unsupervised 3D Point Cloud Learning}
\label{subsec:2.3}
All the above deep neural networks can classify and segment the point cloud data well. However, considering the complexity of labelling 3D data, it is difficult to get enough data with expert labelling for supervised training in many scenarios. Therefore, it is meaningful for exploiting unsupervised or self-supervised learning for 3D point cloud data. PointContrast \cite{xie2020pointcontrast} proposed an Unsupervised framework with U-net as the backbone network. And it demonstrated the transferability of representation learning to 3D point cloud data and the performance enhancement of pre-training to downstream tasks. Lu et al. \cite{lu2018unsupervised,lu20193d} attempted to address skeleton learning on point cloud sequence data. Jiang et al. \cite{jiang2021unsupervised} introduced a simple yet effective unsupervised learning method on point cloud that only considers rotation as the transformation. Info3D \cite{sanghi2020info3d} proposed to extend the InfoMax \cite{velivckovic2018deep} and contrastive learning principles on 3D shapes. It maximized the mutual information between 3D objects and their ``chunks'' to improve the representation in the aligned dataset. FoldingNet \cite{yang2018foldingnet} proposed an autoencoder with graph pooling and MLP layers using the folding operation to deform 2D grids into object surfaces.  However, in 3D medical point cloud data, unsupervised methods are still in great demand. We propose an unsupervised representation learning method, which shows excellent performance for the classification and segmentation of point cloud based intracranial aneurysms.

\section{Method}
\label{sec:3}
\begin{figure*}[ht]
\centering
\includegraphics[width=0.95\linewidth]{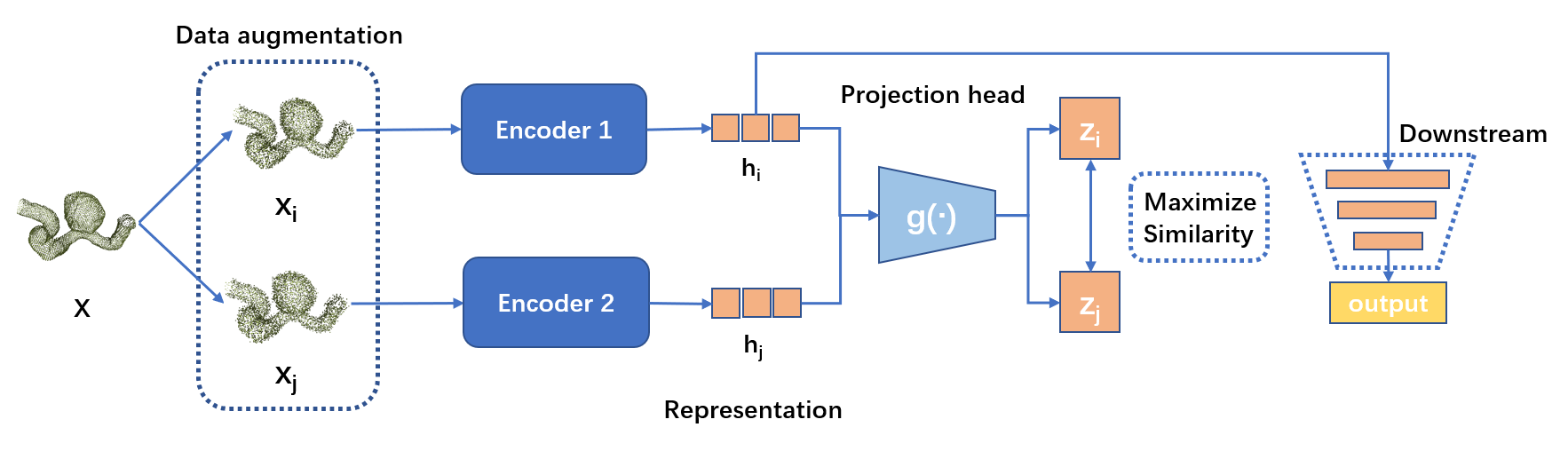}
\caption{The architecture of our method including data augmentation, encoder, projection head, loss function and downstream tasks. We first jitter a point cloud $x$ to construct a pair ($x_{i}$, $x_{j}$). The representation vectors $z_{i}$ and $z_{j}$ to the pair of point clouds are then extracted via the dual-branch encoders and mapping head, and network optimisation is performed by a contrastive loss.The representation $h$ obtained by dual-branch encoders will be used for downstream tasks.
}
\label{architecture} 
\end{figure*}
Our method consists of two stages which are unsupervised learning and downstream tasks. In stage 1, we first perform augmentation on each point cloud to get a pair of augmented samples which are different in pose (Section \ref{subsec:3.1}). 
We then get two representations of a pair of data in a high-dimensional space by means of the dual-branch encoders, which enables each branch of the encoder to extract distinct  features. Next, we map representations to a low-dimensional vector \cite{hadsell2006dimensionality} with a projection head to improve network training speed (Section \ref{subsec:3.2}). Last, we employ a contrastive loss to encourage the representations of the pair of point clouds output by the encoders to be similar in the high-dimensional space (Section \ref{subsec:3.3}). In stage 2, the trained model is used to output unsupervised representations for the downstream task (Section \ref{subsec:3.4}). The downstream task will evaluate the effectiveness of unsupervised learning. Figure \ref{architecture} presents the architecture of our method.

\subsection{Data Augmentation}
\label{subsec:3.1}
We use data augmentation to generate different samples for each point cloud. To generate a pair of data, we consider using data augmentation methods, including jittered, perturbation, and rotation transformations. After experiments, it was found that  \textit{jittered} as data augmentation in both branches gave the best results in the downstream tasks, indicating a more discriminative representation learned by the upstream network. Ablation experiments will be presented in Section \ref{subsec:4.4}.

We take a batch of point clouds with mini-batch size $N$ and input them into the data augmentation module. As shown in the top of Figure \ref{architecture}, for a sample in the mini-batch, we use the jittered function to obtain a pair of samples: one is the jittered point cloud $x_{i}$ and the other is the jittered point cloud $x_{j}$. 
In this way, we have a batch size of $2N$ in this mini-batch. We randomly select $\{x_i,x_j\}$ as a positive pair, and the other $N-1$ pairs, which consist of one of positive sample and one of the other samples, are regarded as negative samples in this mini-batch. 

\begin{figure*}[ht]
\centering
\includegraphics[width=0.95\linewidth]{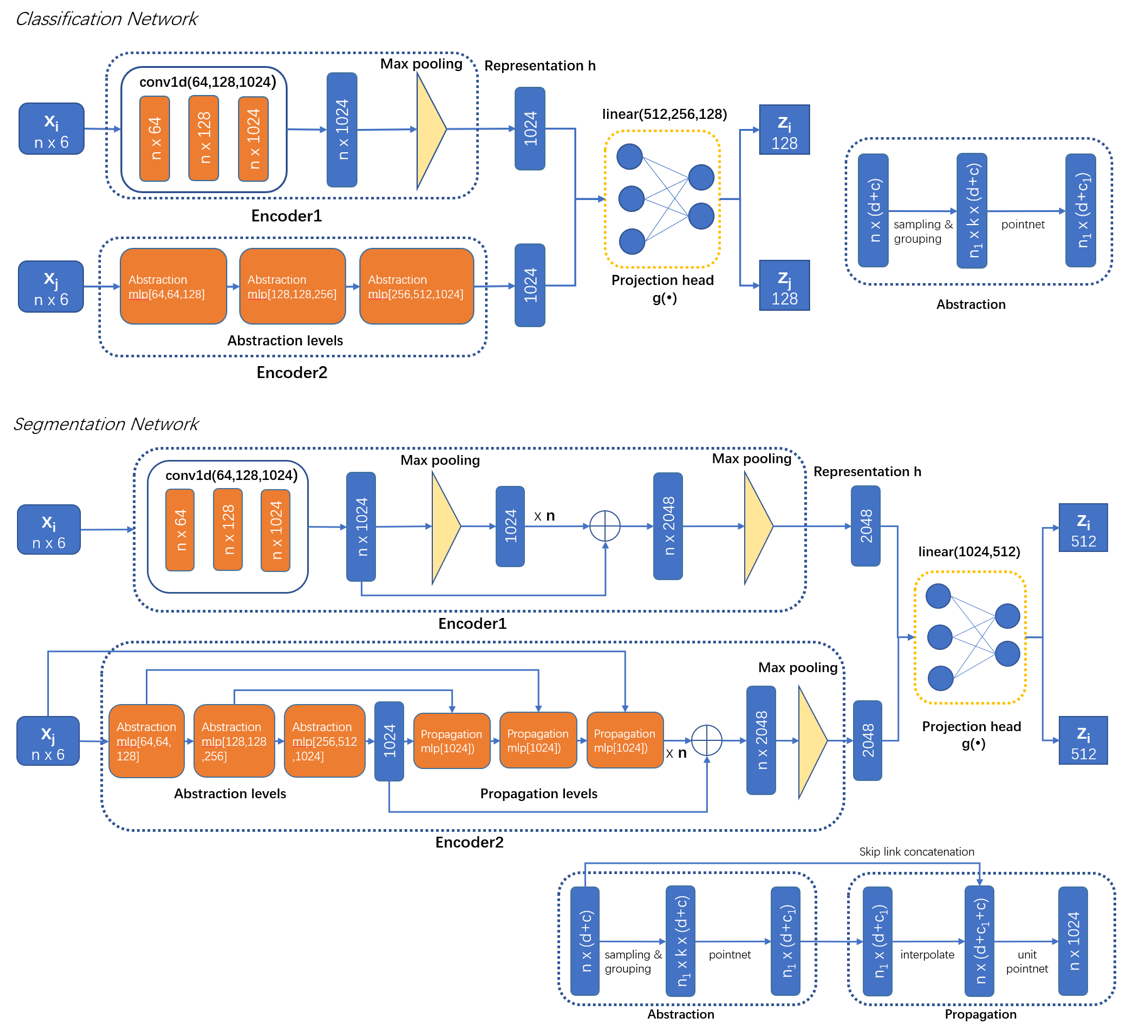}
\caption{Dual-branch Encoders and projection head.  ``conv1d'': 1D convolution, ``linear'': fully connected layers, ``mlp'' stand for multi-layer perceptron. The numbers in brackets represent the layer sizes. All convolutions and fully connected layers include batchnorm and ELU. 
}
\label{encoder} 
\end{figure*}
\subsection{Dual-branch Encoders and Projection Head}
\label{subsec:3.2}
As shown in the left bottom of Figure \ref{encoder}, each pair of samples need to be passed through dual-branch encoders $f(\cdot)$ to obtain two representations $h_i$ and $h_j$. Features are respectively extracted from a pair of data using two different encoders.  Experimentally, we have also compared two different encoders with a common encoder. Ablation experiments will be presented in Section \ref{subsec:4.4}. 

Two encoders are PointNet \cite{qi2017pointnet} and PointNet++ \cite{qi2017pointnet++}, respectively.  The reason for choosing PointNet and PointNet++ is that PointNet can extract global features while PointNet++ can extract local features. This design highlights distinctions in features and allows for more distinctive representation.

\textbf{Classification.}  The first encoder utilizes three consecutive 1D convolutional layers and a max-pooling layer to obtain the representation vector $h$ ($1024$-dimensional). The second encoder consists of three abstraction levels. Each layer abstracts and processes the point set to create a new point set with fewer elements. The input to the abstraction layer consists of an $n\times (d + c)$ matrix formed by n points with $d$-dim coordinates and $c$-dim point features. $n$ is the number of points in a point cloud sample. It outputs a point set group of size $n_{1}\times k\times (d + c)$ by sampling $n_{1}$ centroids and grouping them, where each group corresponds to a local region. $k$ is the number of points sampled from the centroid point's neighbourhood. The subsequent pointnet  layer outputs a local region feature vector $n_{1}\times (d + c_{1})$. We take all the sampled points as a group in the last abstraction layer and output the representation vector $h$ ($1024$-dimensional). We design three linear layers as our projection head $g(\cdot)$ to map each $1024$-dimensional representation vector to a $128$-dimensional vector $z$. 

\textbf{Segmentation.} 
The segmentation encoders are based on the encoders for classification. The $1024$-dimensional representation vector output by encoder 1 is copied $n$ times to form an $n \time 1024$ tensor.  Concatenating it with the $n\times 1024$ tensor obtained from the last convolutional layer gives a $n\times 2048$ tensor. As such, this tensor contains both global features and  features for each point. 
The encoder 2 extends the classification's encoder 2 by adding three propagation layers. We adopt distance-based interpolation and a skip-link across levels propagation strategy. In a propagation level, we propagate point features from $n_{1}\times (d + c_{1})$ points to $n$ points. We achieve feature propagation by concatenating interpolating feature values $c_{1}$ of $n_{1}$ points with skip linked point features from the set abstraction level $c$ of the $n$ points. It outputs a $n\times (d + c_{1}+c)$ vector, which is then passed through the unit pointnet to obtain a $n\times 1024$ tensor. The $1024$-dimensional vector output from the abstraction layer is copied $n$ times and concatenated with the $n\times 1024$-dimensional tensor output from the propagation layer to obtain the final $n\times 2048$ representation tensor.  
The tensor obtained by encoder 1 and encoder 2 are max-pooled separately to obtain a $2048$-dimensional  representation vector. These two vectors are used as the feature representation $h$ for the downstream segmentation network. We design two linear layers as our projection head $g(\cdot)$ to map each $2048$-dimensional representation vector to a $512$-dimensional vector $z$.

\subsection{Contrastive Loss }
\label{subsec:3.3}

We use a contrastive loss function similar to \cite{chen2020simple}. 
With this loss function, unsupervised learning can effectively learn separable features for point clouds. 
After the projection head $g(\cdot)$, for each sample in the mini-batch, we obtain the projection representation $z$. 
For the pair $x_{i}$ and $x_{j}$, we use their projection representations $z_{i}$ and $z_{j}$ to measure the cosine similarity between the two samples, as follows: 

\begin{equation}\label{eq:similarity}
s_{i,j}=\frac{z{_{i}}^{\top }z_{j}}{(\left \| z_{i} \right \|\left \| z_{j} \right \|)}
\end{equation}

Intuitively, the similarity for a positive sample pair should be high. A combination pair of a positive and a negative sample should be low. Then, we use  to get a similar probability of each positive sample pair in a mini-batch. %($S(i,j)$)
$\mathbb{1}_{[k\neq i]}\in\{0,1\}$ is an indicator function evaluating to 1 iff $k\neq i$. The equation for calculating the probability of similarity is as follows:

\begin{equation}\label{eq:softmax}
S(i,j)=\frac{exp(s_{i,j})}{\sum_{k=1}^{2N}\mathbb{1}_{k!=i}exp(s_{i,k})}
\end{equation}

We use the negative logarithm to calculate the loss of the sample pair. This loss has been used in previous works \cite{sohn2016improved,wu2018unsupervised,oord2018representation}. $\tau$ denotes a temperature parameter which scales the input and expands the range of cosine similarity. This loss is known as the normalized temperature-scaled cross-entropy loss \cite{bachman2019learning,oord2018representation} as follows:

\begin{equation}\label{eq:NCE}
l(i,j)=-log\frac{exp(s_{i,j}/\tau )}{\sum_{k=1}^{2N}\mathbb{1}_{k!=i}exp(s_{i,k}/\tau )}
\end{equation}

We calculate the average loss of both $(i,j)$ and $(j,i)$ in the mini-batch. Based on this loss, the representation of the encoder and projection head improves over time, and the trained network places similar samples closer in the representation space. Specifically, the loss function is given by:

\begin{equation}\label{eq:NT-Xent}
L=\frac{1}{2N}\sum_{k=1}^{N}[l(2k-1,2k)+l(2k,2k-1)]
\end{equation}

\subsection{Downstream Tasks}\label{subsec:3.4}
 
We design two simple downstream networks to evaluate the unsupervised learned representations for classification and segmentation, respectively. 
Each point cloud of a vessel segment is fed into the unsupervised dual-branch encoders to obtain two representations. We then concatenate the two representations into one and use this representation as input to train the downstream network.   As for the binary classification task, we use four linear layers (512, 256, 128, 2) as the downstream network. Regarding the segmentation task, we employ four 1D convolutional layers (1024, 512, 256, m), where $m$ is the number of segmentation labels.

\section{Evaluation}\label{sec:4}
\subsection{Datasets}\label{subsec:4.1}

\begin{table*}[ht]
\begin{center}
\newcommand{\tabincell}[2]{\begin{tabular}{@{}#1@{}}#2\end{tabular}}
\begin{tabular}{|l| c c c c c |}
\hline
\multirow{20}*{Supervised}  &  Network    & \#.Points   & V.(\%) & A.(\%) & F1-Score \\[3pt]\hline\hline

~ & SpiderCNN\cite{xu2018spidercnn}   & \tabincell{c}{512 \\ 1024 \\2048} 
                                & \tabincell{c}{98.05 \\ 97.28 \\97.82}  
                                & \tabincell{c}{84.58 \\ 87.9 \\84.89}  
                                & \tabincell{c}{0.8692 \\ 0.8722 \\0.8662}  \\\cline{2-6}
~ & SO-Net\cite{li2018so-net}  & \tabincell{c}{512 \\ 1024 \\2048} 
            & \tabincell{c}{98.76 \\ 98.88 \\98.88}  
            & \tabincell{c}{84.24 \\ 81.21 \\83.94}  
            & \tabincell{c}{0.8840 \\ 0.8684 \\0.8850} \\\cline{2-6}
 ~ & PointCNN\cite{li2018pointcnn}    & \tabincell{c}{512 \\ 1024 \\2048} 
            & \tabincell{c}{98.38 \\ 98.79 \\\textbf{98.95}}  
            & \tabincell{c}{78.25 \\ 81.28 \\85.81}  
            & \tabincell{c}{0.8494 \\ 0.8748 \\\textbf{0.9044}} \\\cline{2-6}
 ~ & DGCNN\cite{wang2019DGCNN}  & \tabincell{c}{512/10 \\ 1024/20 \\2048/40} 
            & \tabincell{c}{95.22 \\ 95.34 \\97.93}  
            & \tabincell{c}{60.73 \\ 72.21 \\83.40}  
            & \tabincell{c}{0.6578 \\ 0.7376 \\0.8594} \\\cline{2-6}
 ~ & PointNet++\cite{qi2017pointnet++}  & \tabincell{c}{512 \\ 1024 \\2048}  
            & \tabincell{c}{98.52 \\ 98.52 \\98.76}  
            & \tabincell{c}{86.69 \\ 88.51 \\87.31}  
            & \tabincell{c}{0.8928 \\ 0.9029 \\0.9016}  \\\cline{2-6}          
 ~ & PointNet\cite{qi2017pointnet}    & \tabincell{c}{512 \\ 1024 \\2048} 
            & \tabincell{c}{94.45 \\ 94.98 \\93.74}  
            & \tabincell{c}{67.66 \\ 64.96 \\69.50}  
            & \tabincell{c}{0.6909 \\ 0.6835 \\0.6916} \\\hline
\multirow{10}*{Unsupervised}& FoldingNet\cite{yang2018foldingnet}                                                   & \tabincell{c}{512 \\ 1024 \\2048} 
                        & \tabincell{c}{91.37 \\ 91.83 \\91.64}  
                        & \tabincell{c}{77.41 \\ 78.28 \\79.54}  
                        & \tabincell{c}{0.6159 \\ 0.6241 \\0.6316} \\\cline{2-6}
 ~ & Our(single\_PN)  & \tabincell{c}{512 \\ 1024 \\2048} 
            & \tabincell{c}{94.33 \\ 94.21 \\94.84}  
            & \tabincell{c}{75.55 \\ 78.33 \\77.05}  
            & \tabincell{c}{0.7233 \\ 0.7424 \\0.7408} \\\cline{2-6}
 ~ & Our(single\_PN++)  & \tabincell{c}{512 \\ 1024 \\2048} 
            & \tabincell{c}{95.33 \\ 95.63 \\95.74}  
            & \tabincell{c}{80.55 \\ 83.60 \\83.41}  
            & \tabincell{c}{0.7679 \\ 0.7968 \\0.7988} \\\cline{2-6}
 ~ & Our(dual)& \tabincell{c}{512 \\ 1024 \\2048} 
            & \tabincell{c}{96.74 \\ 97.45 \\95.41}  
            & \tabincell{c}{82.35 \\ 84.28 \\\textbf{89.47}}  
            & \tabincell{c}{0.8296 \\ 0.8613 \\0.8226} \\\hline  
\end{tabular}
\end{center}
\caption{Classification results of each method. The additional input $K$ is required for DGCNN. PN: PointNet, PN++: PointNet++. }
\label{Classificationtable}
\end{table*}
\textbf{IntrA} \cite{yang2020intra} consists of complete models of aneurysms, generated vessel segments and annotated aneurysm segments. IntrA collected 103 3D models of the entire cerebral vasculature by reconstructing 2D MRA images scanned for patients. IntrA generated 1,909 vessel segments from the complete model, including 1,694 healthy vessel segments and 215 aneurysmal segments. Additionally, 116  aneurysm segments were manually annotated for each point. In IntrA, each sample was represented as a 3D point cloud. Each point $p$ is a 6D vector composing of its coordinates and normal vector. Following IntrA, we combined the generated vessel segments and manually annotated aneurysms to achieve a total of 2,025 samples. These 2,025 vessel segments will be used as the dataset for our unsupervised training. All 2,025 vessel segments will be used for the downstream classification task. 116 annotated aneurysm segments will be used for the downstream segmentation task. 

\textbf{ModelNet-40} \cite{wu20153d} is a collection of 40 categories and 12,311 models culled from the mesh surfaces of CAD models. Following previous practice, 9,843 models are employed for training, and 2,468 for testing. Each point cloud has 2,048  points, and all of the points'  coordinates are normalised to the unit sphere. Each point is a 6D vector made up of its coordinates and normal vector. We take 1,024 points from each item and augment the data by jittered. This dataset is used for comparisons of our method with other unsupervised methods. 

\subsection{Experimental Setting}\label{subsec:4.2}

For unsupervised training, we use the Adam optimizer with a weight decay of $10^{-6}$.  The mini-batch size is set to 32.  The number of epochs is 200. The initial learning rate is $10^{-3}$. The learning rate is scheduled to be multiplied by 0.5 in every 10 epochs. We use jittered as the data augmentation method, which directly adds Guassian noise to every coordinate and normal information of input point clouds. In the encoder 2, the number of points sampled from the centroid point's neighbourhood $k$ is set to [32, 64, ``None'']. ``None'' means that all points are sampled. The projection head outputs a feature  $z$,  the dimension of which is set to 128 for the classification task and 512  for the segmentation task. In the loss function, the temperature parameter $\tau$ is set to 0.5. 

For the downstream network, the optimizer, the number of epochs, representation dimensions, initial learning rate, learning rate decay schedule  and mini-batch size are the same as those in unsupervised training. We sample 512,1024 and 2048 points separately in each point cloud for both experiments. For classification task weight decay is set to $10^{-6}$ and size of linear is set to [512, 256, 128, a]. $a$ is the number of categories in the point cloud. For segmentation task, weight decay is set to 1.0 and size of MLPs is set to [1024, 512, 256, b].  $b$ is the number of categories of points in the point cloud.

Experiments were implemented using PyTorch on a GeForce GTX 1080 GPU. For  IntrA dataset, the time for both unsupervised training on classification and segmentation is approximately 1 hour. The downstream classification and  segmentation training are approximately 40 minutes and 50 minutes, respectively.

\subsection{Experimental Results}\label{subsec:4.3}

We evaluate the classification task and the segmentation task separately on \textbf{IntrA}  \cite{yang2020intra}.To demonstrate the generalisation of our method, we also perform the classification task on \textbf{ModelNet-40} \cite{wu20153d}, and then compare our method with start-of-the-art unsupervised methods to verify the effectiveness of our method. 

\textbf{Classification task.} On IntrA, we evaluate the performance using three metrics: (1) V. Accuracy, measuring the percentage of correctly predicted healthy vessels' samples over all healthy vessels' samples, (2) A. Accuracy, indicating the percentage of correctly predicted aneurysm vessels' samples over all aneurysm vessels' samples, (3) F1 score, representing the harmonic average of precision and recall and evaluating the quality of the model. On ModelNet-40, we evaluate the performance using overall accuracy. 
 
As shown in Table \ref{Classificationtable}, as for our dual-branch encoders method with PointNet and PointNet++ backbones (i.e. PN, PN++), 1,024 sample points have the best results in terms of the F1 score and V. Accuracy compared with other numbers of sample points. The results for 512 sample points are still impressive, though the number of points in each point cloud is much smaller. Although the 2,048 sample point result is not the best in terms of F1 score and V. accuracy. 
\textit{Notice that our method in 2,048 sample point has the best A. Accuracy results compared with all mentioned methods. The ability to identify aneurysms is essential in this case.}
Furthermore, we find that the A. accuracy increase with more  sample points. Compared with other supervised methods, our results are better than the supervised PointNet in all metrics. Besides, it also outperforms more advanced supervised networks such as DGCNN in general. Result of 1,024 sample points is very close to SO-Net and SpiderCNN in terms of F1-Score.  We also compare our method with FoldingNet, one of the most representative unsupervised methods.  Obviously, our method performs better on all metrics. The effectiveness of unsupervised learning is inherently limited due to the unsupervised nature, and the tubular structure of intracranial aneurysms is much less prominent compared with other data. It causes our method (dual) to perform less well than supervised PN++. 

\begin{table}[ht]
\begin{center}
\begin{tabular}{|l|c|}
\hline
Method & ModelNet40(\%)\\
\hline\hline
SPH\cite{kazhdan2003rotation} & 68.2\% \\
LFD\cite{chen2003visual} & 75.5\%  \\
T-L Network\cite{girdhar2016learning} & 74.4\%  \\
VConv-DAE\cite{sharma2016vconv} & 75.5\%  \\
3D-GAN\cite{wu2016learning} & 83.3\%  \\
Latent-GAN\cite{achlioptas2017representation} & 85.7\% \\
FoldingNet\cite{yang2018foldingnet} & 88.4\%\\
PointCapsNet\cite{zhao20193d} & 88.9\% \\
MultiTask\cite{hassani2019unsupervised} & 89.1\%  \\
\hline
Our(dual) & \textbf{90.79}\% \\
\hline
\end{tabular}
\end{center}
\caption{Classification accuracy of unsupervised learning on ModelNet40. }
\label{ModelNet40}
\end{table}

As shown in Table \ref{ModelNet40}, we  compare the performance of our model with other unsupervised methods on ModelNet40 \cite{wu20153d}. we can see that our method outperforms all other unsupervised methods, which again confirms its effectiveness in unsupervised representation learning. 

\begin{table}[ht]
\begin{center}
\newcommand{\tabincell}[2]{\begin{tabular}{@{}#1@{}}#2\end{tabular}}
\begin{tabular}{|l c c c|}
\hline
Network    & \#.Points & IoU\_V.(\%) & IoU\_A.(\%) \\\hline\hline
SO-Net      & \tabincell{c}{512 \\ 1024 \\2048}  
            & \tabincell{c}{94.22 \\ 94.42 \\\textbf{94.46}}  
            & \tabincell{c}{80.14 \\ 80.99 \\\textbf{81.40}} \\\hline
PN++        & \tabincell{c}{512 \\ 1024 \\2048} 
            & \tabincell{c}{93.42 \\93.35 \\93.24}  
            & \tabincell{c}{76.22 \\ 76.38 \\76.21}\\\hline            
PointCNN    & \tabincell{c}{512 \\ 1024 \\2048} 
            & \tabincell{c}{92.49 \\ 93.47 \\93.59}  
            & \tabincell{c}{70.65 \\ 74.11 \\73.58} \\\hline
SpiderCNN   & \tabincell{c}{512 \\ 1024 \\2048} 
            & \tabincell{c}{90.16 \\ 87.95 \\87.02}  
            & \tabincell{c}{67.25 \\ 61.60 \\58.32}  \\\hline
PointGrid   & \tabincell{c}{16/2 \\ 16/4 \\32/2} 
            & \tabincell{c}{78.32 \\ 79.49 \\80.11}  
            & \tabincell{c}{35.82 \\ 38.23 \\42.42} \\\hline
PointNet    & \tabincell{c}{512 \\ 1024 \\2048} 
            & \tabincell{c}{73.99 \\ 75.23 \\74.22}  
            & \tabincell{c}{37.30 \\ 37.07 \\37.75} \\\hline 
Our(PN)   & \tabincell{c}{512 \\ 1024 \\2048} 
            & \tabincell{c}{80.05 \\ 82.54 \\81.65}  
            & \tabincell{c}{44.66 \\ 46.55 \\48.45}\\\hline
Our(PN++)   & \tabincell{c}{512 \\ 1024 \\2048} 
            & \tabincell{c}{80.05 \\ 82.05 \\82.65}  
            & \tabincell{c}{40.66 \\ 41.42 \\42.45}\\\hline
Our(dual)   & \tabincell{c}{512 \\ 1024 \\2048} 
            & \tabincell{c}{82.25 \\ 84.35 \\82.65}  
            & \tabincell{c}{48.66 \\ 50.92 \\51.45}\\\hline
\end{tabular}
\end{center}
\caption{Segmentation results of each network. }
\label{Segmentationtable}
\end{table}
\textbf{Segmentation task.} Following \cite{yang2020intra}, we evaluate the segmentation performance using two metrics: (1) V. IoU, indicating the IoU of heathly vessel, and (2) A. IoU, indicating the IoU of aneurysm vessel.

As shown in Table \ref{Segmentationtable}, as for our method (dual), 1,024 sample points have the best results in terms of V. IoU. But 2,048 sample points have the best results in terms of A. IoU.  In comparison, our method outperforms the supervised PointNet on both V. IoU and A. IoU. \textit{Notice that our method is better than more advanced supervised networks like PointGrid}. 
Compared to the supervised PointNet, which is trained with only 116 labelled samples, our method is able to learn unsupervised features from a much wider range of data, thus facilitating downstream network training. Our method generally generates better results with increasing the point number, and produces better results than the supervised PointNet in both metrics. Our method (PN++) is still inferior to the supervised PointNet++, which is considered to be limited by the unsupervised nature. 

\subsection{Ablation Studies}\label{subsec:4.4}
We explore the factors that make our method  effective through ablation experiments. We conduct two ablation experiments to further understand which data augmentation is more effective, and the effect of dual-branch encoders. We also analyse the effectiveness of our method in the case of sparse labels. The ablation experiments sample 1,024 points in each point cloud. 

\textbf{Augmentation.} We try to find the best data augmentation method for our unsupervised method, by considering three different augmentation methods including \textit{rotation, jittered and perturbation.} 

\begin{table}[ht]
\begin{center}
\begin{tabular}{|l|c|c|c|}
\hline
Augmentation & V.(\%) & A.(\%) & F1-Score\\\hline\hline
rotation & 95.23 & 75.52 & 0.7637\\
perturbation & 95.35 & 81.87 & 0.8121\\
jittered \& perturbation & 95.63 & 81.76 & 0.8240\\
jittered & \textbf{97.45} & \textbf{84.28} & \textbf{0.8613}\\
\hline
\end{tabular}
\end{center}
\caption{Ablation study on augmentation. }
\label{Augmentation}
\end{table}

Rotation means randomly rotating the point cloud along the $Y$-axis. Perturbation means randomly rotating the point cloud by a small angle along the $XYZ$-axis. Jittered is the addition of Gaussian noise to the $XYZ$ coordinates and normal information of the point cloud. As shown in Table \ref{Augmentation}, jittered is the best data augmentation for both branches, and the method with both jittered and perturbation is the second and generally better than the perturbation for both branches. The data augmentation with both branches rotation is the least effective. Based on the results, we can find that the data augmentation using jittered allows the encoder to learn the distinctive features of the point cloud more effectively, thereby giving better results. 

\begin{figure}[htbp]
\centering
\begin{minipage}[b]{0.55\linewidth}
\includegraphics[width=1\linewidth]{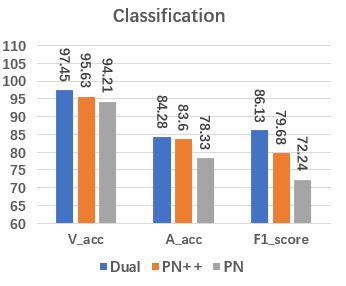}
\label{1}
\end{minipage}%
\begin{minipage}[b]{0.45\linewidth}
\includegraphics[width=1\linewidth]{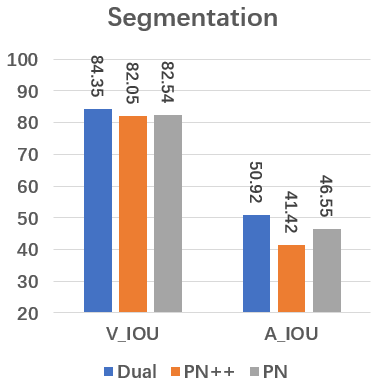}
\label{2}
\end{minipage}
\caption{Ablation study on dual-branch encoders. }
\label{fig:ablation}
\end{figure}

\textbf{Dual-branch encoders.} In order to investigate the effectiveness of dual-branch encoders, experiments are designed to compare it with the traditional single encoder method. As shown in Figure \ref{fig:ablation}, the dual-branch encoders method has the best performance for both classification and segmentation tasks, in particular for the classification task. Besides, the single encoder method based on the more advanced PN++ is inferior to the PN-based method. This is probably due to contrastive learning depends largely on global information. 

Based on these results, we have the following findings: 
\begin{itemize}
    \item Dual-branch encoders are able to extract more discriminative features.  In particular, the two encoders in our design are PN and PN++ respectively where PN focuses on global features and PN++ on local features. 
    \item Contrastive learning can better understand the distinctions between the features extracted by the two encoders. Therefore it is more effective than a single encoder. In our design, PN and PN++ as encoders can better highlight the distinctions between the local and global features of a point cloud sample,  allowing the contrastive loss to optimise the network more effectively.
    \item Contrastive learning is excellent at describing objects as a whole, but is weak at describing them at the point scale. Our method achieves outstanding results in classification tasks, but is moderately effective in segmentation tasks. This is because our contrastive learning is not a comparison between points but between point clouds as a whole. 
\end{itemize}

\begin{table}[ht]
\begin{center}
\newcommand{\tabincell}[2]{\begin{tabular}{@{}#1@{}}#2\end{tabular}}
\begin{tabular}{|l|c|c|c|c|}
\hline
Network & Label(\%) & V.(\%) & A.(\%) & F1-Score\\\hline\hline
PointNet & \tabincell{c}{10 \\ 5 \\ 1 }  
        & \tabincell{c}{87.43 \\ 86.53 \\ 84.86} 
        & \tabincell{c}{53.33 \\ 42.85 \\ 30.58} 
        & \tabincell{c}{0.3298 \\ 0.3012 \\ 0.2485} \\\hline
PointNet++ & \tabincell{c}{10 \\ 5 \\ 1 }  
        & \tabincell{c}{94.37 \\ 92.89 \\ 89.55} 
        & \tabincell{c}{70.58\\ 63.23 \\ 45.07} 
        & \tabincell{c}{0.7111 \\ 0.6370 \\ 0.4637} \\\hline
Our(dual)& \tabincell{c}{10 \\ 5 \\ 1 }  
        & \tabincell{c}{95.34 \\ 94.39 \\ 90.84} 
        & \tabincell{c}{71.19 \\ 67.27 \\ 58.31} 
        & \tabincell{c}{0.7294 \\ 0.6935 \\ 0.5712} \\\hline

\hline
\end{tabular}
\end{center}
\caption{Ablation study for limited labeled Data. }
\label{Labeled}
\end{table}

\textbf{Limited Labeled data.} In real-world situations, we frequently lack sufficient labeled data. We divide the original dataset into two parts, A and B, assume A to be the unlabeled data and B to be the labeled data, to represent such circumstance. The percentages of labeled data are set to 10\%, 5\% and 1\%, respectively. In unsupervised learning, we use A+B to pre-train the model, and then use B for training the downstream tasks.  Because of the nature of supervised learning, only B is used for  other supervised training. The experiments are set up with a classification task and performed on the IntrA dataset. As shown in Table \ref{Labeled}, the accuracy of the classification gradually decreases as the amount of annotated data decreased. However, the accuracy of our model consistently outperforms that of the supervised models. This suggests that our method is more robust by making use of the unlabeled data for unsupervised learning. 
An interesting point is to combine evolutionary optimization with the proposed method to enhance the performance on limited labeled data  \cite{nakane2020application}.

\section{Conclusion}\label{sec:5}
% \vspace{0.2cm}
In this work, we have presented an unsupervised representation learning method for the classification and segmentation of 3D intracranial aneurysms. It first augments a point cloud into two samples, and pairs them up for going through the dual-branch encoders and a subsequent common projection head. Distinctive  features are learned by maximising the correspondence for a pair. The representations learned by the unsupervised trained encoders are used as input for the downstream tasks. Experiments demonstrated that our method is effective in learning unsupervised representations and can achieve better or comparable performance than state-of-the-art supervised and unsupervised learning methods.

%%%%%%%%% REFERENCES
{\small
\bibliographystyle{ieee_fullname}
\bibliography{egbib}
}

\end{document}